\documentclass[twocolumn,showpacs,showkeys,preprintnumbers,amsmath,amssymb]{revtex4}
\usepackage{graphicx}% Include figure files
\usepackage{dcolumn}% Align table columns on decimal point
\usepackage{bm}% bold math

\begin{document}

%\preprint{APS/123-QED}

\title{SPONTANEOUSLY APPEARING DISCRETE MOVING KINKS IN NONLINEAR
ACOUSTIC CHAIN WITH REALISTIC POTENTIALS}
% Force line breaks with \\

\author{L.S.Metlov}
\author{Yu.V.Eremeichenkova}
\email{metlov@ukr.net}
\affiliation{Donetsk Physico-Technical
Institute, Ukrainian Academy of Sciences, \\83114, R.Luxemburg
str. 72, Donetsk, Ukraine}

\date{\today}% It is always \today, today,

\begin{abstract}
Molecular dynamic simulations are performed to investigate a
long-time evolution of different type initial signals in nonlinear
acoustic chains with realistic Exp-6 potential and with power
ones. Finite number of long-lifetime kink-shaped excitations is
observed in the system in thermodynamic equilibrium. Dynamical
equilibrium  between the processes of their growth and decay is
found.
\end{abstract}


%Use showkeys class option if keyword
\maketitle

\section{\label{sec:level1}Introduction}

Power expansions of interatomic potentials (such as 2-4 and 2-3-4
potentials) are frequently used for investigation of strongly
nonlinear phenomena \cite{b1,b2,b3,b4,b5,b6}. However, for
soliton-like excitations with large amplitudes one can not be sure
that high-order anharmonisms may be omitted. So, realistic (e.g.
Lennard-Jones, Morse) potentials were used to take all-order
anharmonisms into account \cite{b6,b7,b8}.

Several types of soliton-like excitations were found in discrete
monoatomic chains with anharmonic intersite potentials (acoustic
chains).

Breathers - time-periodic space-localized
one-parameter$^{\footnotemark [1]}$ \footnotetext [1] {Necessary
conditions for existence of two-parameter moving breathers
(frequency and velocity are a parameters) and numerical algorithm
for their searching are proposed in \cite{b9}} modes with
frequency as a parameter are revealed in FPU chains with 2-4
potential \cite{b1,b10}. Another solutions - dark solitons (phase
shift in stable time-periodic extended nonlinear mode) are proved
to exist in FPU chains \cite{b11,b12}. In monoatomic acoustic
chains with realistic Lennad-Jones and Morse potentials breathers
are not observed in numerical experiments, and, moreover, they are
forbidden \cite{b6}. However, they arise if time-periodic driving
field is applied to the chain \cite{b7,b8}.

Moving kinks - time non-periodic  solitons with step-shaped
profiles of atom displacements \cite{b15,b16,b17,b18,b19} and
pulses - solitons with bell-shaped profiles \cite{b17,b20,b21} are
predicted in acoustic chains. These solitons are one-parameter
ones with a velocity as a parameter. Numerical results for pulses
were obtained in \cite{b20,b21} for FPU and Toda chains.  Exact
non-topological moving kinks are well known in integrable acoustic
Toda chain \cite{b15}. Existence theorems of non-topological
moving kinks are also proved for non-integrable FPU chains
\cite{b16,b17} and for acoustic chains with interatomic potentials
of arbitrary powers \cite{b18}. We don't know other solutions for
moving kinks in discrete case for acoustic chains. In continual
(long wave) limit  moving kinks are observed in real ultrasonic
experiment and investigated numerically taking third-order
anharmonism into account (KDV equation) \cite{b19}.

Thus, no any soliton-like solutions have been obtained in discrete
case for non-integrable acoustic chains with realistic interatomic
potentials. In this paper spontaneous creation of discrete moving
kinks  is observed in numerical experiment for the chain with such
potential.

The aim of this article is molecular dynamics study of discrete
acoustic chain with realistic interatomic potential at high
temperatures after thermalization.  The role of different orders
of anharmonicity in power expansion of realistic potential is
investigated.

The methods used are described in the Sec.II. Numerical results
and discussions for thermalized gas of spontaneously appearing
moving kinks are given in the Sec.III.

\section{\label{sec:level1} Methods and approximations}

Equations of motion

\begin{equation}\label{1}
    m \ddot{r_{i}}=\frac{1}{2} \partial/\partial r_{i}
    \sum_{ij}V(|r_{i}-r_{j}|)
\end{equation}
are integrated numerically with periodic boundary conditions for
acoustic chain containing 100 atoms. Nearest neighbor
approximation is used. Analogous results are obtained when six
neighbors and 1500 atoms are taken into account.

We use the system of units in which energy is measured in [K],
mass in [a.m.u.] does, and distance is measured in the units of
equilibrium interatomic distance [$d_{0}$]. The unit of time is
equal to the period of harmonic zone boundary phonon mode
($T_{0}=8.3716\cdot10^{-13}$ sec). It corresponds to the case of
neon with atom mass $m$=20.18 a.m.u. Velocities are measured in
the units of sound velocity $v_{s}=12.72 [(K/a.m.u.)^{1/2}]$.

Six-order symplectic Yoshida algorithm \cite{b22} is applied for
numerical simulations. Time step is equal to 0.008 $T_{0}$ which
provides energy conservation with the accuracy $\triangle
E/E=10^{-6}$ during whole simulation time ($t=10^{6} T_{0}$).

Realistic interatomic Exp-6 potential

\begin{eqnarray}\label{2}
\nonumber V^{exp-6}(|r_{i}-r_{j}|) =
    A_{0} \exp(-\alpha(x_{ij}-1)) - \frac{\alpha A_{0}}{6} x_{ij}^{-6};\\
    x_{ij}=|r_{i}-r_{j}|/d_{0}
\end{eqnarray}
is used with the parameters $A_{0}$=35.9335 K, $\alpha$=13.6519
obtained $ab-initio$ for the dimer of neon \cite{b23}. Single
empirical parameter $d_{0}=3.091 \dot{A}$ is fitted to equilibrium
interatomic distance in neon dimer \cite{b24}.

Four types of initial conditions are used.
\begin{itemize}
    \item Zone boundary mode with wave vector $k=\pi$ ($\pi$-mode). Equal opposite
initial velocities $|V_{0}|=0.24 v_{s}$ were assigned to neighbor
atoms.
    \item White noise. Random initial velocities were fixed.
    \item Shock waves. Initial velocities $|V_{0}|=0.96 v_{s}$ directed inside
the chain were given to three atoms at each end of the chain.
    \item Exact breathers. Initial displacements of atoms were fixed
according to breather's exact form $A(\ldots , 0, -1/2, 1, 1/2, 0,
\ldots )$ for odd-parity breather and $A(\ldots , 0, -1, 1, 0,
\ldots )$ for even-parity one \cite{b3}. The amplitude is A=0.279
$d_{0}$.
\end{itemize}

In all the cases the same energy $E$=90 K/atom is fed into the
system by the initial conditions. This energy is two times larger
then cohesive energy of neon dimer.

To investigate the role of different order anharmonisms, the
simulations have performed on acoustic chains with power
interatomic potentials. These potentials are defined expanding
Exp-6 potential

\begin{widetext}
\begin{equation}\label{3}
    V^{exp-6}(|r_{i}-r_{j}|) =    V_{0} +
    \frac{K_{2}}{2} (x_{ij}- 1)^{2} +
    \frac{K_{3}}{3} (x_{ij}- 1)^{3} +
    \frac{K_{4}}{4} (x_{ij}- 1)^{4} + \ldots
\end{equation}
\end{widetext}
and taking higher powers into account consequently.

\section{\label{sec:level1} Results and discussions}

Excluding the cases specially mentioned, acoustic chain with Exp-6
potential is considered  with different initial conditions in all
the subsections. Chains with power and Toda potentials are also
reported.

\subsection{\label{sec:level2}$\pi$-mode}

To the time $\thickapprox 50 T_{0}$ initial $\pi$-mode perturbed
by numerical noise is destroyed by period-doubling instability
(see also \cite{b6}). In the following, the motion of atoms become
more and more chaotic. Simultaneously, the process of space
concentration of energy takes place. Well-localized solitons with
the width compatible to lattice constant are spontaneously created
since $\thickapprox 600 T_{0}$. Thus, the solitons in the chain
with realistic potential appear after thermalization. On the
contrary, breathers in FPU chain arise directly as a result of
$\pi$-mode destruction, and the system is termalized after decay
of breathers.

The thermalization is defined here as a state when the energy
distribution of atoms obtained in numerical simulation is close to
Gibbs one. The system persists in such state for a long time. The
deviations from Gibbs distribution caused by solitons are possible
at high-energy tile only.

\begin{figure*}
\includegraphics [width=7in] {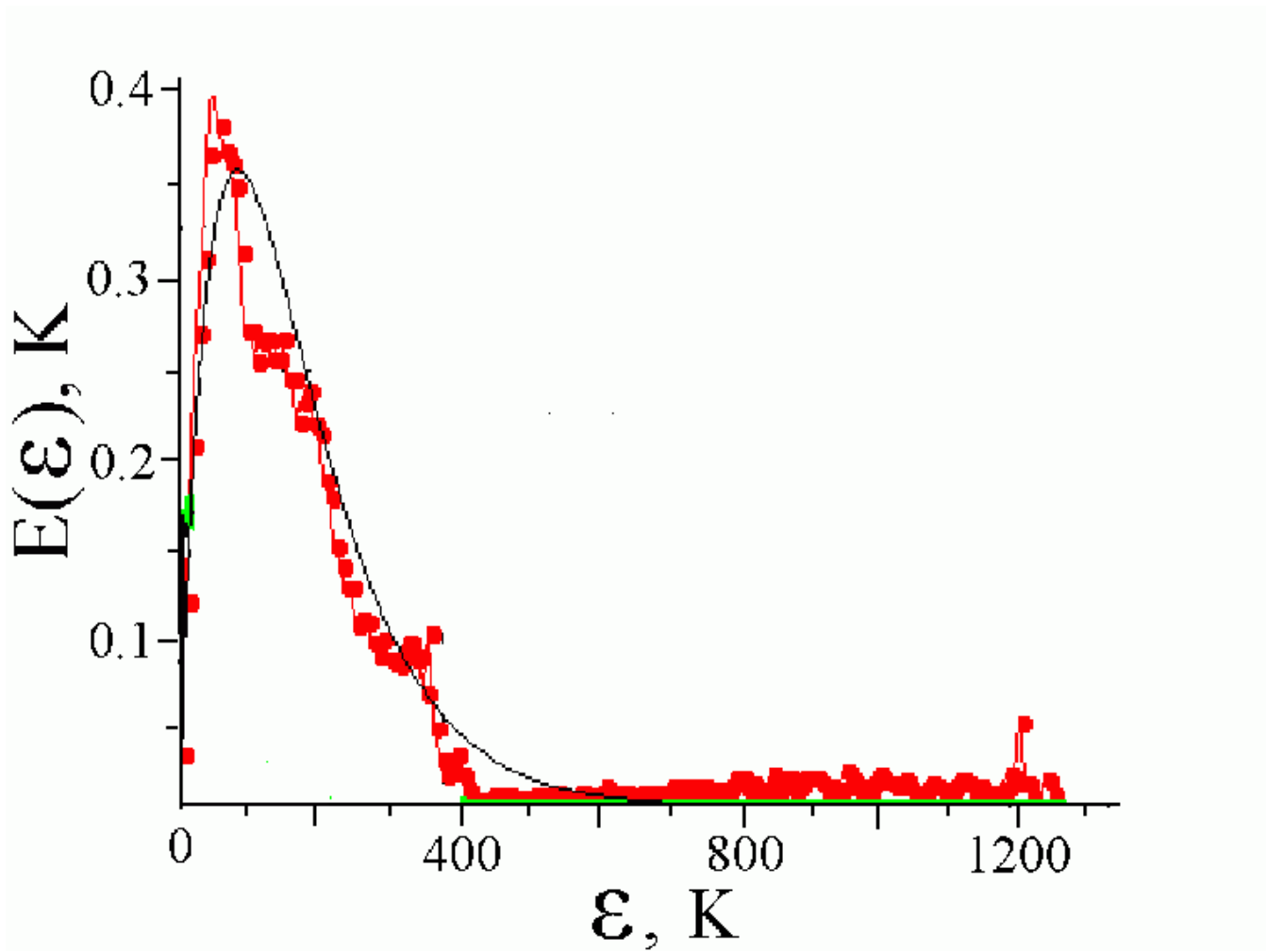}
\caption{\label{f1a} Energy distribution of atoms
$E(\varepsilon)=\varepsilon n(\varepsilon)$. Solid curve
corresponds to Gibbs form of the $n(\varepsilon)$, the circles
denote the distribution $E(\varepsilon)$ obtained in numerical
experiment. To get "experimental" distribution the energies of
atoms are averaged for the time 11 $T_{0}$. The distribution
corresponds to one of the most intensive kinks ($t=188947
T_{0}$).}
\end{figure*}

In the Fig.~\ref{f1a} energy distribution of atoms
$E(\varepsilon)=\varepsilon n(\varepsilon)$ is given. The
$E(\varepsilon)$ is defined so that
$dE(\varepsilon)=E(\varepsilon) d \varepsilon$ is total energy of
atoms which fall into energy interval $(\varepsilon, \varepsilon +
d \varepsilon)$. The $n(\varepsilon) d \varepsilon$ is the number
of atoms with the energies residing in the same interval. Solid
curve $E(\varepsilon)$ corresponds to Gibbs distribution
$n(\varepsilon)$ obtained in harmonic approximation. The circles
denote the distribution $E(\varepsilon)$ observed in numerical
experiment, when one of the most intensive solitons appears. This
soliton gives a local peak on the tile of "experimental"
$E(\varepsilon)$. The position of the peak on energy scale
manifests that the energy concentrated in the soliton is
$\thickapprox 15$ times as mush as average energy of an atom in
the system. At all the rest energies "experimental" distribution
agrees with Gibbs one. It indicates that majority of atoms move
chaotically.

Fractional deviation of "experimental" distribution
$n(\varepsilon)$ from Gibbs one is reported in Fig.~\ref{f1b}
(numbered by 1).
\texttt{\begin{figure}
\includegraphics [width=3.89in] {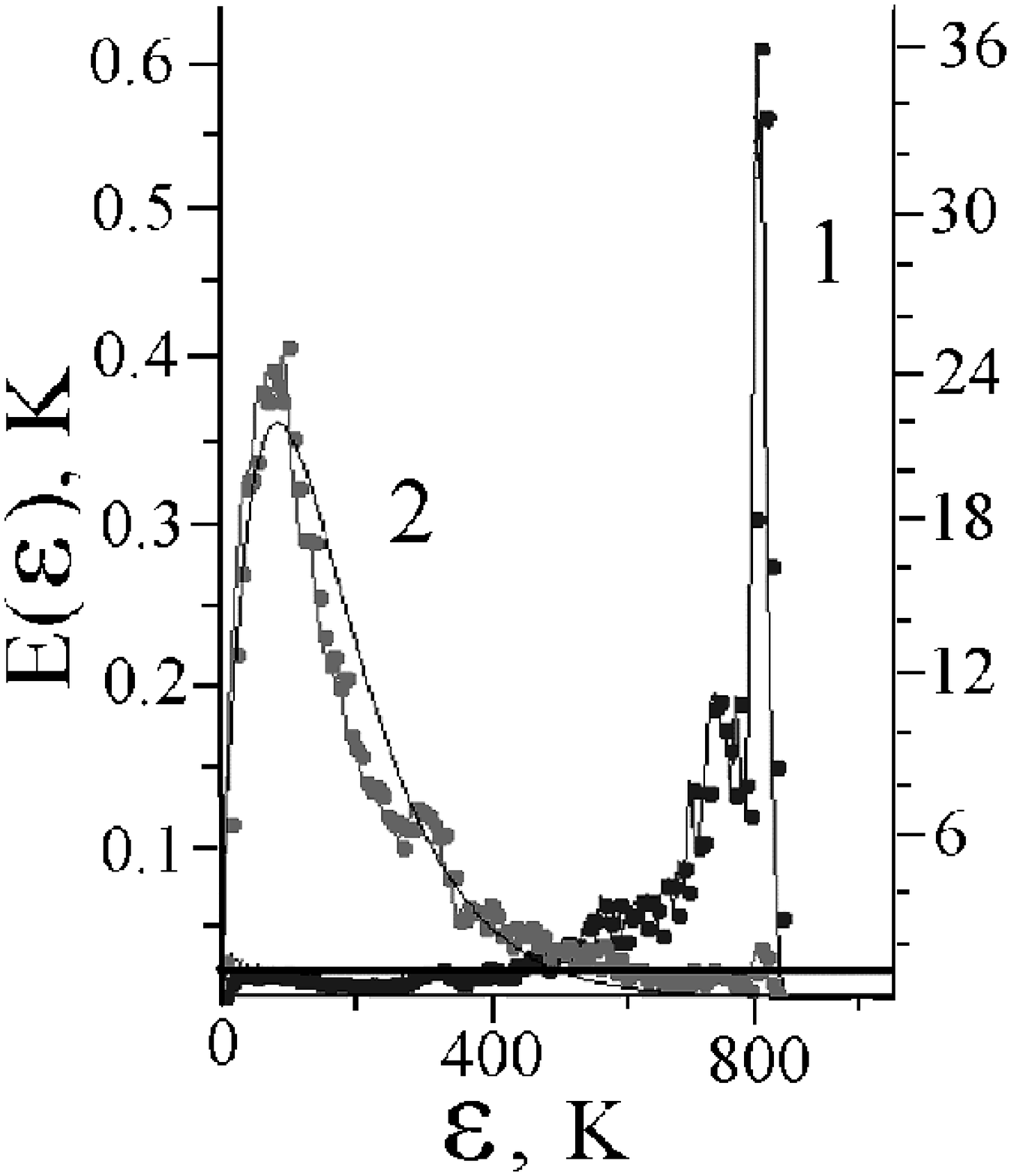}
\caption{\label{f1b} Fractional deviation of "experimental"
distribution $n(\varepsilon)$ from Gibbs one,
$n_{exper}(\varepsilon)/n_{Gib}(\varepsilon)$, at $t=2316248
T_{0}$ (1 - black sircles connected by the line; right scale). The
digit 2 denotes "experimental" and Gibbs distributions
$E(\varepsilon)$ at the same $t$.}
\end{figure}}
"Experimental" and Gibbs distributions $E(\varepsilon)$ are also
given here (numbered by 2). The value of "experimental"
$n(\varepsilon)$ is $\thickapprox 36$ times as much as Gibbs one
near the peak at the tile, while $n_{nexper}/n_{Gib} \thicksim 1$
at all the rest energies.

\begin{figure*}
\includegraphics [width=6.5in] {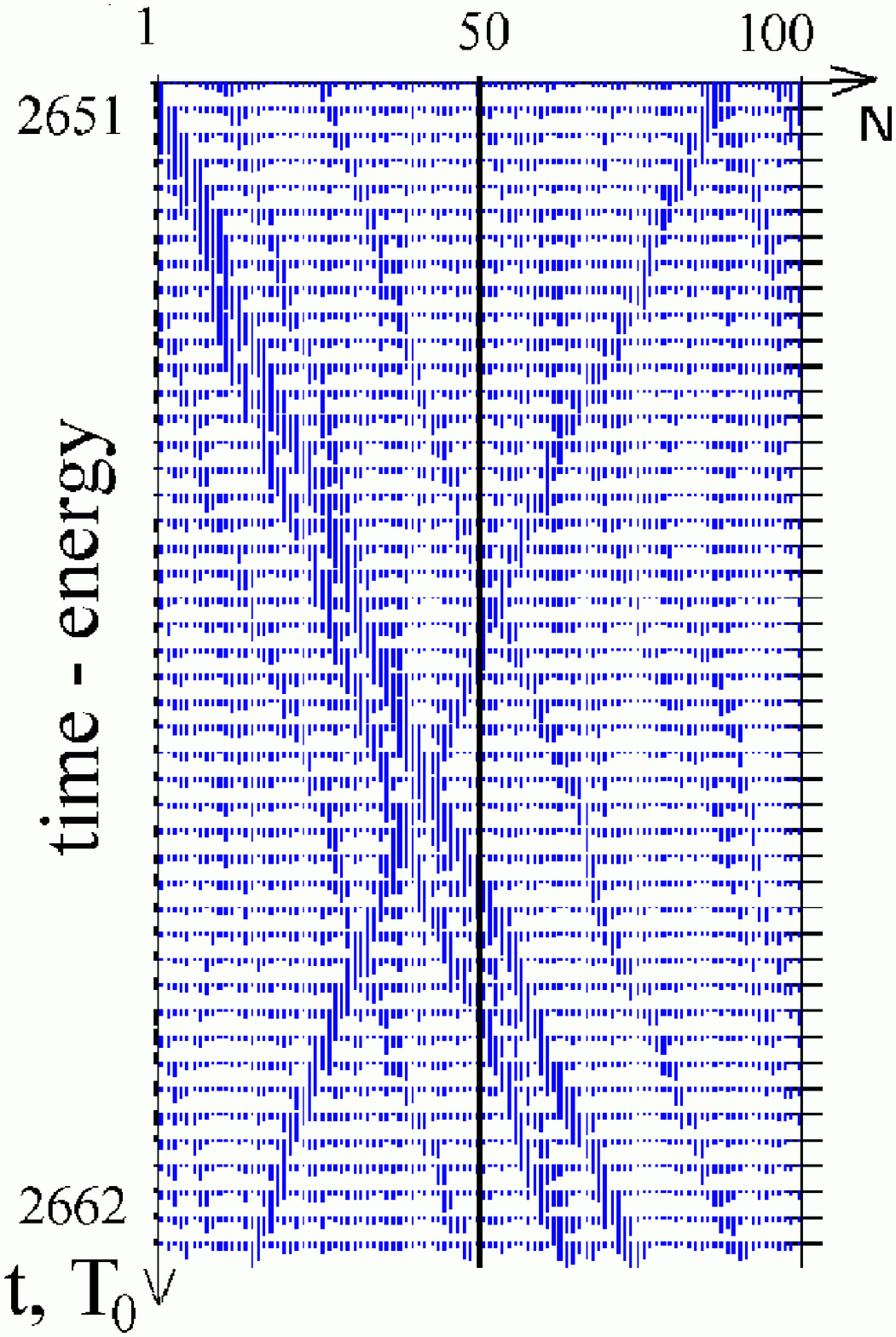}
\caption{\label{f3} Interaction of moving kinks.}
\end{figure*}

Typical pattern of map of  tracks of solitons is plotted in
Fig.~\ref{f2a} where single powerful soliton is seen clearly.
\texttt {\begin{figure}
\includegraphics [width=4.3in] {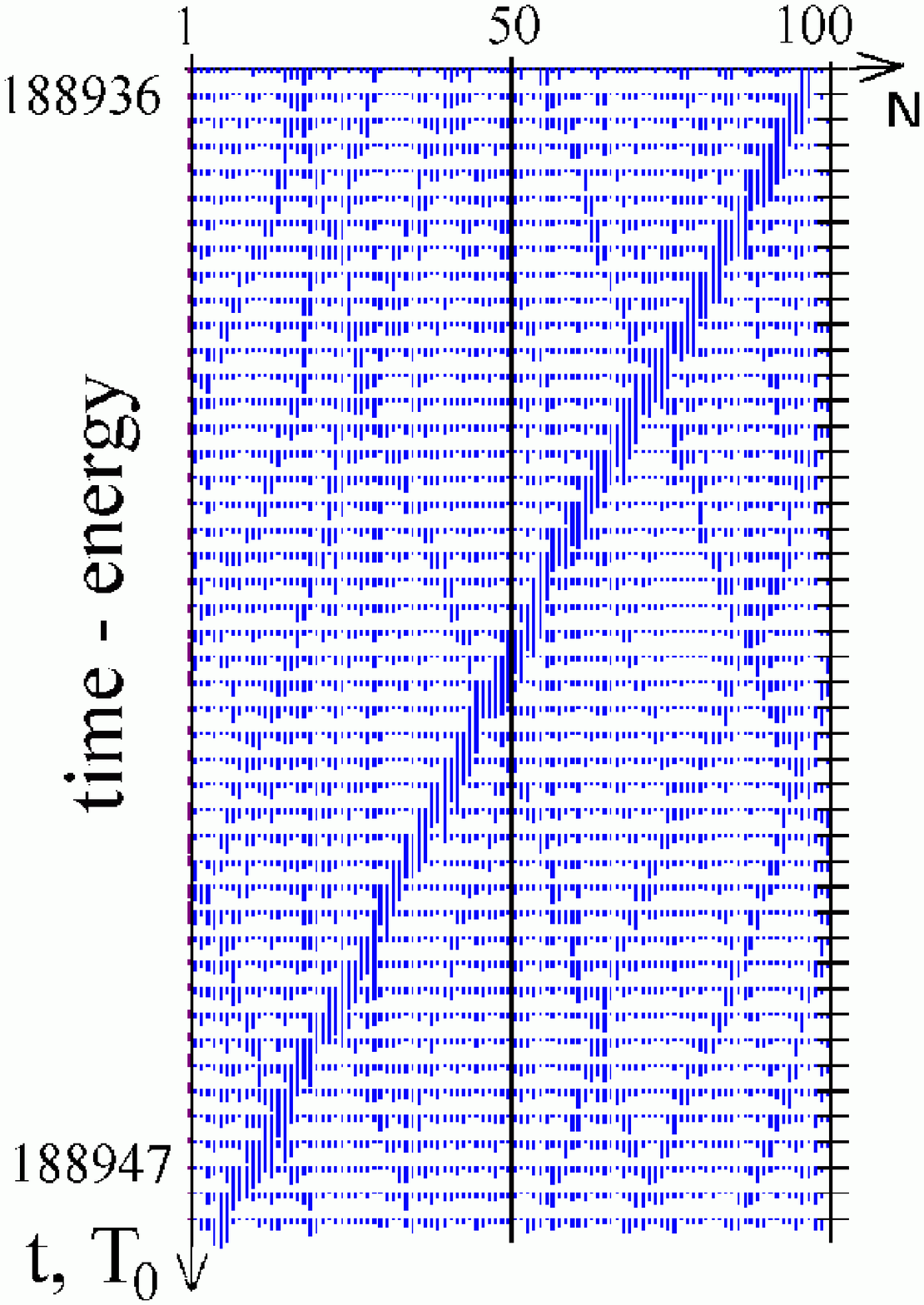}
\caption{\label{f2a} Map of tracks of moving kink from numerical
experiment of the Fig.~\ref{f1a}. Horizontal axis indicates the
position along the chain, vertical axis corresponds to the time
(time is going downward). The energies and velocities of atoms are
denoted as short vertical lines with the length proportional to
the magnitude (positive direction is downward).}
\end{figure}}
Time periodicity in the motion of atoms is not observed. Velocity
of the soliton (the slope of the curve of energy maximum position
via the time) is equal to $v=2.6 v_{s}$. Numerical experiment
shows that the the velocities and the amplitudes of observed
solitons are connected uniquely. Therefore, these solitons are
one-parameter ones. Besides, they are pass each through others
without loss of their individuality like to solitons in integrable
systems (Fig.~\ref{f3}).

The form of obtained solitons  can be determined from enlarged
fragment of the map of tracks plotted in Fig.~\ref{f2b}.
\texttt{\begin{figure}
\includegraphics [width=2.7in] {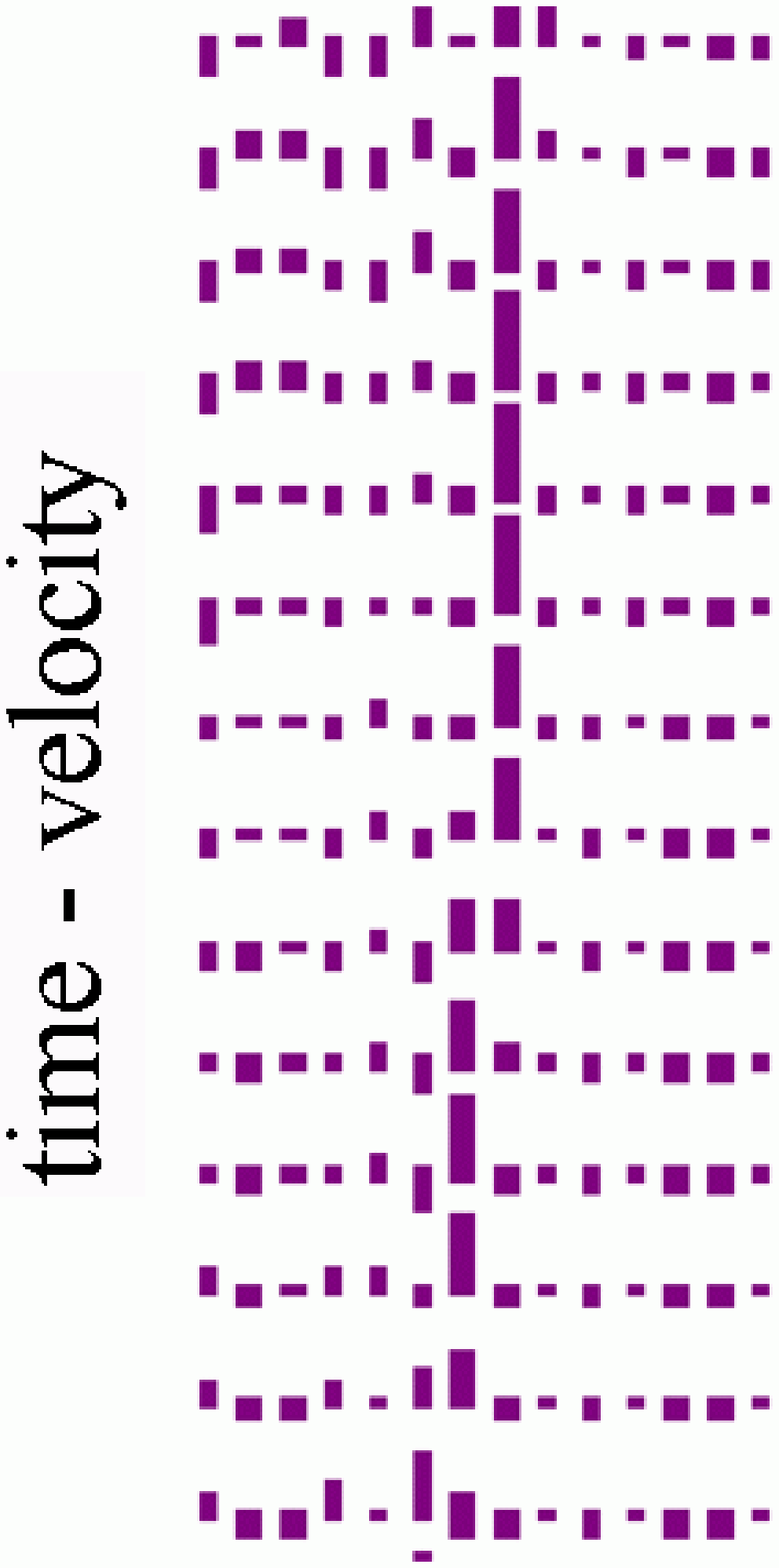}
\caption{\label{f2b}The form of moving kink from the
Fig.~\ref{f2a}. Time step is equal 0.022 $T_{0}$.}
\end{figure}}The velocities of atoms are given here. The atom in the soliton
moves in the same direction that the soliton does and transfer its
momentum to the next neighbor. It can be said with good accuracy
that this soliton is localized on two sites only. Space dependence
of atoms velocity has bell-shaped form at each moment of time.
Therefore, the displacements of atoms have step-shaped form, and
the solitons can be identified as non-topological moving kinks.

\begin{figure*}
\includegraphics{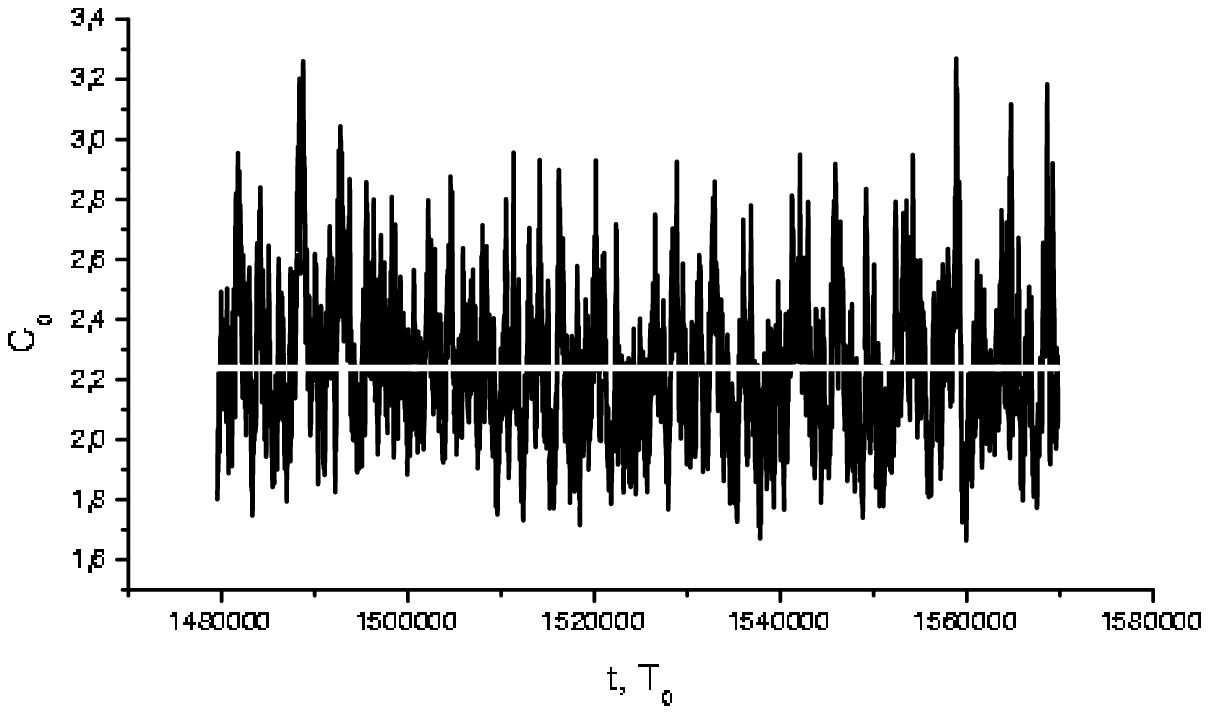}
\caption{\label{f4} Parameter of energy concentration via the time
(black line). White line is average value $<C_{0}>$.}
\end{figure*}

Let use the parameter

\begin{equation}
    C_{0}=N \sum_{i=1}^{N} E_{i}^{2}\diagup (\sum_{i=1}^{N} E_{i})^{2}
\end{equation}
as quantitative indicator of energy concentration \cite{b4}. The
$C_{0}$=1 if all the atoms have the same energies, $C_{0}$=1.75
for equilibrium Gibbs distribution in harmonic approximation
\cite{b4}, and $C_{0}=N$ if all the energy is concentrated on one
site only. Parameter $C_{0}$ indicates energy concentration in all
the solitons existing simultaneously, but speaks nothing about the
number of solitons and their individual contributions. In
Fig.~\ref{f4} the $C_{0}$ via the time is given for the
chain after thermalization. For the time interval under
consideration $C_{0}$ oscillates around its average value in the
frames 1.7-3.3, and it doesn't exceed 1.5-4.2 during whole
simulation time ($t=10^{6} T_{0}$). It follows from this
oscillating behavior of the $C_{0}$ that the probabilities of
kink's growing and decay are equal.

Different periodic components of $C_{0}(t)$ are separated by fast
Fourier transform which is performed taking 8192 time points into
account. To suppress weak high-frequency fluctuations of the
$C_{0}$ caused by thermal motion of atoms initial time dependence
of $C_{0}$ was smoothed by adjacent averaging using 25 time
points. Amplitude Fourier spectrum smoothed using 5 frequency
points is plotted in Fig.~\ref{f5} (the curve 1). A number of
\texttt{\begin{figure}
\includegraphics [width=3.7in] {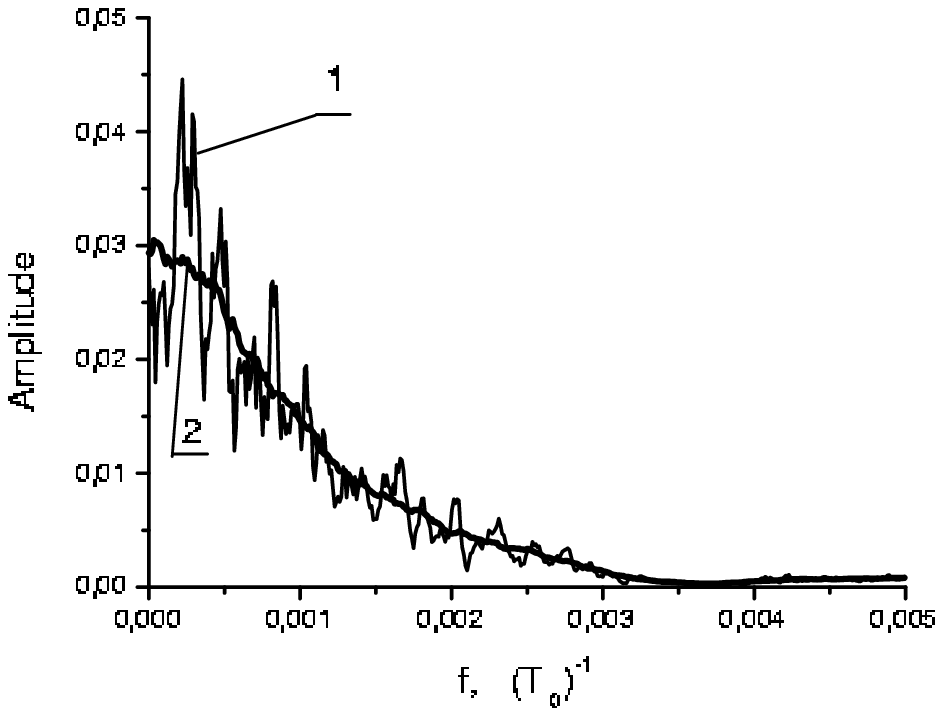}
\caption{\label{f5} Amplitude Fourier spectrum of $C_{0}$(t). The
curve 1 is the spectrum smoothed using 5 points; the curve 2 is
the spectrum smoothed using 50 points. }
\end{figure}} considerable peaks is observed at the frequencies $10^{-5}-10^{-3}
(T_{0})^{-1}$. However, the width of these peaks is compatible to
the distance between them, and each peak by itself is seen to be
meaningless. Thus, one can evaluate maximal frequency (minimal
period) of long-time oscillations of $C_{0}(t)$ only. For
convenience we smooth frequency dependence of Fourier amplitudes
using more number of points (50 points, the curve 2 in the
Fig.~\ref{f5}). We define maximal frequency as a frequency at
which Fourier amplitudes drop two times. Then, minimal period of
$C_{0}(t)$ oscillations is evaluated to be $\thicksim 1000 T_{0}$,
which agrees with the lifetime of most intensive kinks defined
directly using tracks maps.

Numerical simulations show a number of kinks to exist
simultaneously in the chain. Some of them persist on the stage of
growing and others are on the stage of decay. Moreover, there are
a long-time fluctuations of number of kinks which correlate with
the fluctuations of energy concentration parameter $C_{0}$: the
more the magnitude of $C_{0}$ the less the number of
kinks$^{\footnotemark [2]}$ \footnotetext [2] {Similar correlation
was observed on acoustic chain with Lennard-Jones potential
\cite{b25}.}. For example, for strong concentration of energy
($C_{0}$=3.7) corresponding to the Fig.~\ref{f2a} single kink runs
along the chain. The number of kinks remains, practically,
unchanged when enlarged amounts of energy of 300 K/atom and 1000
K/atom are initially fed into the chain, while the energy
concentrated in each kink enhances.

In conclusion, there is an hierarchy of excitations in the chain
after thermalization, starting with most strong well-localized and
long-lifetime kinks and finishing with phonons. Weak excitations
deserve as the background with which strong kinks interact. Time
after time strong kinks appear spontaneously from the background.
For certain time each kink exhibits systematic tendency to growth,
which changes by stable tendency to decay. The behavior of
ensemble of such kinks leads to oscillating time dependence of
energy concentration parameter, and all the properties averaged
for a long enough time are steady.

\subsection{\label{sec:level2} White noise}

Although there is a stage of chaotic motion of atoms after decay
of initial mode the question remains about effect of initial
conditions on the form of created solitons. To exclude this effect
we use white noise as initial conditions. The system will choose
most appropriate solitons.

At the beginning of the simulations atoms move chaotically, but
the distribution $E(\varepsilon)$ is differ a bit from Gibbs one.
The process of energy concentration starts immediately. After
$\thickapprox 80 T_{0}$ the $E(\varepsilon)$ agrees with Gibbs
one. Well localized kinks are created since $\thickapprox 400
T_{0}$. Their form and properties are identical to ones for the
kinks arising after $\pi$-mode decay.

Thus, whatever initial conditions, acoustic chain with realistic
Exp-6 potential comes to the same state characterized by finite
number of long-lifetime high-energy supersonic moving kinks.

\subsection{\label{sec:level2} Shock waves}

The form of spontaneously appearing moving kinks is similar to the
form of shock waves. We adopt shock waves as initial conditions,
which permit to set the amplitudes of the kinks in controlled
manner.

Instantly each initial shock wave including three atoms splits
into three kink-like excitations moving with the velocities 4.5-5
$v_{s}$; $C_{0}$=17.4. They are localized on two sites only.
Kink-like excitations lose the energy slowly. To $\thickapprox
1100 T_{0}$ they become slightly delocalized in space: two, three
or more atoms moving in the same direction are contained in the
kinks. The parameter $C_{0}$ drops to 3.1, kinks velocities
decrease to 2.3-2.8 $v_{s}$. So, quantitative characteristics of
shock waves become compatible to ones for spontaneously created
kinks.

Numerical experiments on shock waves with different initial
amplitudes show that the more energy is stored in the wave the
more its lifetime.

We consider the propagation of the same shock waves in integrable
Toda chain where non-topological moving kinks are exact solutions.
Like FPU chain, each initial shock wave splits into three kinks
with the form, velocities and parameter $C_{0}$ similar to ones
for Exp-6 potential. In the following these kinks transfer a bit
of energy to background atoms, and, further, neither the form nor
the energies of the kinks are altered to the end of simulation
(t=28000 $T_{0}$).

\subsection{\label{sec:level2} Exact breathers}

It was found that in monoatomic acoustic chains with realistic
Lennad-Jones and Morse potentials breathers are not created
\cite{b6}. We have performed numerical simulations on the chain
with Exp-6 potential with initial displacements of atoms in the
form of single odd-parity breather placed in the center of the
chain. For the time less then $T_{0}$ the breather decays into two
kinks with equal energies moving in opposite directions. In the
following shock wave scenario is realized. Kink-like excitations
are delocalized slightly in space, and to 5000 $T_{0}$ the
parameter $C_{0}$ and kinks velocities decrease and become
compatible to the characteristics of spontaneously created kinks.
Further, each kink grows and decays as it was described above. The
simulations with even parity breather give analogous result.

Instability of breathers in the chain with Exp-6 potential is
caused by strong asymmetry of the potential. Inside the breather,
one of interatomic bonds is compressed and its potential energy is
increased. If the potential would be symmetrical, increasing of
the potential energy would be equal when the atoms both approach
each other and remove. So, if the slope of potential curve is
large enough, the pair of atoms would vibrates almost like
isolated bound state as it does in FPU chain. In Exp-6 potential
attractive part is much weaker then repulsive one, and the atoms,
if approached, remove almost freely. They collide with neighbor
atoms and transfer whole momentum to them. As a result, two kinks
run along the chain in opposite directions.

\subsection{\label{sec:level2} Power potentials}

The form and some properties of moving kinks in the chain with
Exp-6 potential are similar to ones of the solutions of Bussinesk
equation. It can be obtained from Eq.~\ref{1} in continual
approximation if cubic anharmonism only is taken into account.
This similarity would suggest that high order anharmonisms could
be neglected. To clarify the role of different order anharmonisms
we perform the simulations for the chains with the potentials
obtained as power expansion Eq.~\ref{3} of Exp-6 potential.

The next cases are considered (initial conditions - $\pi$-mode).
\begin{itemize}
    \item 2-3 potential. $\pi$-mode decays, the solitons appear and
collapse immediately.
    \item 2-4 and 2-3-4 potentials. $\pi$-mode decays by modulational
instability, and the breathers occur in agreement with
\cite{b1,b2,b3,b4,b5,b6}. Maximal velocity of breather is
$\thicksim 0.1 v_{s}$, the lifetime is $\thickapprox 120 T_{0}$
(2-4 potential) and $\thickapprox 50 T_{0}$ (2-3-4 potential).
Unlike the kinks, breathers interact each with other strongly.
After breathers decay movement of atoms becomes chaotic, energy
distribution agrees with Gibbs one. In the following kink-like
solitons are spontaneously created in the chain$^{\footnotemark
[3]}$ \footnotetext [3] {Numerical simulation shows kink-like
solitons to arise in the chains with 2-4 and 2-3-4 potentials
regardless of initial conditions.}. However, lifetime of these
kinks is very short ($\thickapprox 50 T_{0}$), indicating that
both kinks and breathers interact strongly in the chains with 2-4
and 2-3-4 potentials.
    \item 2-3-4-5-6 and 2-3-4-5-6-7-8 potentials. $\pi$-mode is
stable.
    \item 2-3-4-5-6-7-8-9-10 potential. $\pi$-mode decays trough
period-doubling instability. Solitons identical to kinks in the
chain with Exp-6 potential are created (with long lifetime and
weakly interacted each with others).
\end{itemize}

Thus, in the chain with 2-3 potential nonlinear term is not
balanced by dispersion one (at least at the energy 90 K/atom), and
solitons are collapse. Cubic anharmonism added to 2-4 potential
give only quantitative changes in the picture of solitons
formation. The 5-6 and 7-8 powers stabilize $\pi$-mode. Higher
(9-10) powers give the change of the type of instability, and the
properties of kinks coincide with ones in realistic Exp-6
potential. So, quantitatively small corrections give qualitative
change of atoms behavior, and realistic interatomic potentials
should be used to avoid such difficulties in real crystals.

\section{\label{sec:level1}Summary}

The behavior of nonlinear acoustic chain with realistic Exp-6
potential is investigated. Regardless of the type of initial
conditions, moving non-topological kink-shaped solitons occur
spontaneously after thermalization of the system. These kinks are
one-parameter solitons with the velocity as a parameter. They
interact each with others weakly and pass each through others
without loss of their individuality like solitons in integrable
systems. Traveling along the chain, each kink, firstly, collect
the energy from the background atoms, and, in the following,
transfer it to small-amplitude phonons. Dynamical equilibrium
between the processes of kink's growth and decay takes place. So,
there is finite number of high-energy excitations in the system at
each moment of time. Time intervals when the most amount of energy
is concentrated in the kinks are repeated periodically. Average
lifetime of a kink is estimated to be $\thicksim 1000 T_{0}$.

Acoustic chains with power interatomic potentials are investigated
adding higher powers of atom displacements consequently up to ten
powers. In the chains with 2-4 and 2-3-4 potentials discrete
breathers are observed in agreement with \cite{b1,b2,b3,b4,b5,b6}.
After breathers decay thermal equilibrium is achieved and
kink-like solitons also appear in these chains. However, these
kinks interact strongly each with others and have short lifetime
(50-120 $T_{0}$). Thus, unlike breathers resided to power
potentials only, the kinks can be created both in power and
realistic potentials whatever initial conditions.

Molecular dynamics study of $\pi$-mode stability in the chains
with power potentials shows the change of the type of instability
(from modulational to period-doubling) when high (9 and 10) powers
are added. It indicates that extreme caution is required using
power expansions of interatomic potentials in real crystals.

Thus, in thermal equilibrium long-lifetime high-energy supersonic
moving kinks appear in acoustic chain with Exp-6 potential at high
temperatures. One can expect that similar excitations can exist in
three-dimensional case and give considerable enhancing of thermal
conductivity at high temperatures and also effect on long-time
strength of solids.

%\newpage
\bibliography{Metlov_sol}% Produces the bibliography via BibTeX.

\begin{thebibliography}{23}
\expandafter\ifx\csname natexlab\endcsname\relax\def\natexlab#1{#1}\fi
\expandafter\ifx\csname bibnamefont\endcsname\relax
  \def\bibnamefont#1{#1}\fi
\expandafter\ifx\csname bibfnamefont\endcsname\relax
  \def\bibfnamefont#1{#1}\fi
\expandafter\ifx\csname citenamefont\endcsname\relax
  \def\citenamefont#1{#1}\fi
\expandafter\ifx\csname url\endcsname\relax
  \def\url#1{\texttt{#1}}\fi
\expandafter\ifx\csname urlprefix\endcsname\relax\def\urlprefix{URL }\fi
\providecommand{\bibinfo}[2]{#2}
\providecommand{\eprint}[2][]{\url{#2}}

\bibitem[{\citenamefont{Dolgov}(1986)}]{b1}
\bibinfo{author}{\bibfnamefont{A.}~\bibnamefont{Dolgov}},
  \bibinfo{journal}{Fiz.\ Tv.\ Tela\ (Sov.)} \textbf{\bibinfo{volume}{28}},
  \bibinfo{pages}{1641} (\bibinfo{year}{1986}).

\bibitem[{\citenamefont{Burlakov et~al.}(1990)\citenamefont{Burlakov, Kiselev,
  and Rupasov}}]{b2}
\bibinfo{author}{\bibfnamefont{V.}~\bibnamefont{Burlakov}},
  \bibinfo{author}{\bibfnamefont{S.}~\bibnamefont{Kiselev}}, \bibnamefont{and}
  \bibinfo{author}{\bibfnamefont{V.}~\bibnamefont{Rupasov}},
  \bibinfo{journal}{Phys.\ Lett.\ A} \textbf{\bibinfo{volume}{147}},
  \bibinfo{pages}{130} (\bibinfo{year}{1990}).

\bibitem[{\citenamefont{Sandusky et~al.}(1992)\citenamefont{Sandusky, Page, and
  Shmidt}}]{b3}
\bibinfo{author}{\bibfnamefont{K.}~\bibnamefont{Sandusky}},
  \bibinfo{author}{\bibfnamefont{J.}~\bibnamefont{Page}}, \bibnamefont{and}
  \bibinfo{author}{\bibfnamefont{K.}~\bibnamefont{Shmidt}},
  \bibinfo{journal}{Phys.\ Rev.\ B} \textbf{\bibinfo{volume}{46}},
  \bibinfo{pages}{6161} (\bibinfo{year}{1992}).

\bibitem[{\citenamefont{Cretegny et~al.}(1998)\citenamefont{Cretegny, Dauxois,
  Ruffo, and Torchini}}]{b4}
\bibinfo{author}{\bibfnamefont{T.}~\bibnamefont{Cretegny}},
  \bibinfo{author}{\bibfnamefont{T.}~\bibnamefont{Dauxois}},
  \bibinfo{author}{\bibfnamefont{S.}~\bibnamefont{Ruffo}}, \bibnamefont{and}
  \bibinfo{author}{\bibfnamefont{A.}~\bibnamefont{Torchini}},
  \bibinfo{journal}{Physica\ D} \textbf{\bibinfo{volume}{121}},
  \bibinfo{pages}{109} (\bibinfo{year}{1998}).

\bibitem[{\citenamefont{Bichham et~al.}(1993)\citenamefont{Bichham, Kiselev,
  and Sievers}}]{b5}
\bibinfo{author}{\bibfnamefont{S.}~\bibnamefont{Bichham}},
  \bibinfo{author}{\bibfnamefont{S.}~\bibnamefont{Kiselev}}, \bibnamefont{and}
  \bibinfo{author}{\bibfnamefont{A.}~\bibnamefont{Sievers}},
  \bibinfo{journal}{Phys.\ Rev.\ B} \textbf{\bibinfo{volume}{47}},
  \bibinfo{pages}{14206} (\bibinfo{year}{1993}).

\bibitem[{\citenamefont{Sandusky and Page}(1994)}]{b6}
\bibinfo{author}{\bibfnamefont{K.}~\bibnamefont{Sandusky}} \bibnamefont{and}
  \bibinfo{author}{\bibfnamefont{J.}~\bibnamefont{Page}},
  \bibinfo{journal}{Phys.\ Rev.\ B} \textbf{\bibinfo{volume}{50}},
  \bibinfo{pages}{866} (\bibinfo{year}{1994}).

\bibitem[{\citenamefont{Rossler and Page}(1995)}]{b7}
\bibinfo{author}{\bibfnamefont{T.}~\bibnamefont{Rossler}} \bibnamefont{and}
  \bibinfo{author}{\bibfnamefont{J.}~\bibnamefont{Page}},
  \bibinfo{journal}{Phys.\ Lett.\ A} \textbf{\bibinfo{volume}{204}},
  \bibinfo{pages}{418} (\bibinfo{year}{1995}).

\bibitem[{\citenamefont{Rossler and Page}(1997)}]{b8}
\bibinfo{author}{\bibfnamefont{T.}~\bibnamefont{Rossler}} \bibnamefont{and}
  \bibinfo{author}{\bibfnamefont{J.}~\bibnamefont{Page}},
  \bibinfo{journal}{Phys.\ Rev.\ Lett} \textbf{\bibinfo{volume}{87}},
  \bibinfo{pages}{1287} (\bibinfo{year}{1997}).

\bibitem[{\citenamefont{Flach and Kladko}(1999)}]{b9}
\bibinfo{author}{\bibfnamefont{S.}~\bibnamefont{Flach}} \bibnamefont{and}
  \bibinfo{author}{\bibfnamefont{K.}~\bibnamefont{Kladko}},
  \bibinfo{journal}{Physica\ D} \textbf{\bibinfo{volume}{127}},
  \bibinfo{pages}{61} (\bibinfo{year}{1999}).

\bibitem[{\citenamefont{Flach and Willis}(1998)}]{b10}
\bibinfo{author}{\bibfnamefont{S.}~\bibnamefont{Flach}} \bibnamefont{and}
  \bibinfo{author}{\bibfnamefont{C.}~\bibnamefont{Willis}},
  \bibinfo{journal}{Phys.\ Rep.} \textbf{\bibinfo{volume}{295}},
  \bibinfo{pages}{181} (\bibinfo{year}{1998}).

\bibitem[{\citenamefont{Chubykalo and Kivshar}(1993)}]{b11}
\bibinfo{author}{\bibfnamefont{O.}~\bibnamefont{Chubykalo}} \bibnamefont{and}
  \bibinfo{author}{\bibfnamefont{Y.}~\bibnamefont{Kivshar}},
  \bibinfo{journal}{Phys.\ Lett.\ A} \textbf{\bibinfo{volume}{178}},
  \bibinfo{pages}{123} (\bibinfo{year}{1993}).

\bibitem[{\citenamefont{Kivshar}(1993)}]{b12}
\bibinfo{author}{\bibfnamefont{Y.}~\bibnamefont{Kivshar}},
  \bibinfo{journal}{Phys.\ Rev.\ Lett.} \textbf{\bibinfo{volume}{70}},
  \bibinfo{pages}{3055} (\bibinfo{year}{1993}).

\bibitem[{\citenamefont{Toda}(1989)}]{b15}
\bibinfo{author}{\bibfnamefont{M.}~\bibnamefont{Toda}},
  \emph{\bibinfo{title}{Theory of nonlinear lattices}}
  (\bibinfo{publisher}{Springer Verlag, Berlin}, \bibinfo{year}{1989}).

\bibitem[{\citenamefont{Carr and B.McLeod}(1997)}]{b16}
\bibinfo{author}{\bibfnamefont{J.}~\bibnamefont{Carr}} \bibnamefont{and}
  \bibinfo{author}{\bibnamefont{B.McLeod}}, \emph{\bibinfo{title}{Solitary
  waves on lattices}}, \bibinfo{howpublished}{Preprint} (\bibinfo{year}{1997}).

\bibitem[{\citenamefont{Friesecke and Wattis}(1994)}]{b17}
\bibinfo{author}{\bibfnamefont{G.}~\bibnamefont{Friesecke}} \bibnamefont{and}
  \bibinfo{author}{\bibfnamefont{J.}~\bibnamefont{Wattis}},
  \bibinfo{journal}{Comm.\ Math.\ Phys.} \textbf{\bibinfo{volume}{161}},
  \bibinfo{pages}{391} (\bibinfo{year}{1994}).

\bibitem[{\citenamefont{Smets and Willem}(1997)}]{b18}
\bibinfo{author}{\bibfnamefont{D.}~\bibnamefont{Smets}} \bibnamefont{and}
  \bibinfo{author}{\bibfnamefont{M.}~\bibnamefont{Willem}},
  \bibinfo{journal}{J.\ Fun.\ Anal.} \textbf{\bibinfo{volume}{149}},
  \bibinfo{pages}{266} (\bibinfo{year}{1997}).

\bibitem[{\citenamefont{Hao and Maris}(2001)}]{b19}
\bibinfo{author}{\bibfnamefont{H.-Y.} \bibnamefont{Hao}} \bibnamefont{and}
  \bibinfo{author}{\bibfnamefont{H.}~\bibnamefont{Maris}},
  \bibinfo{journal}{Phys.\ Rev.\ B.} \textbf{\bibinfo{volume}{64}},
  \bibinfo{pages}{064302} (\bibinfo{year}{2001}).

\bibitem[{\citenamefont{Hochstrasser et~al.}(1989)\citenamefont{Hochstrasser,
  Mertens, and Buttner}}]{b20}
\bibinfo{author}{\bibfnamefont{D.}~\bibnamefont{Hochstrasser}},
  \bibinfo{author}{\bibfnamefont{F.}~\bibnamefont{Mertens}}, \bibnamefont{and}
  \bibinfo{author}{\bibfnamefont{H.}~\bibnamefont{Buttner}},
  \bibinfo{journal}{Physica D} \textbf{\bibinfo{volume}{35}},
  \bibinfo{pages}{259} (\bibinfo{year}{1989}).

\bibitem[{\citenamefont{Eilbeck and Flesh}(1990)}]{b21}
\bibinfo{author}{\bibfnamefont{J.}~\bibnamefont{Eilbeck}} \bibnamefont{and}
  \bibinfo{author}{\bibfnamefont{R.}~\bibnamefont{Flesh}},
  \bibinfo{journal}{Phys.\ Lett.\ A} \textbf{\bibinfo{volume}{149}},
  \bibinfo{pages}{200} (\bibinfo{year}{1990}).

\bibitem[{\citenamefont{Yoshida}(1990)}]{b22}
\bibinfo{author}{\bibfnamefont{H.}~\bibnamefont{Yoshida}},
  \bibinfo{journal}{Phys.\ Lett.\ A} \textbf{\bibinfo{volume}{150}},
  \bibinfo{pages}{262} (\bibinfo{year}{1990}).

\bibitem[{\citenamefont{Eremeichenkova et~al.}()\citenamefont{Eremeichenkova,
  Metlov, and Morozov}}]{b23}
\bibinfo{author}{\bibfnamefont{Y.}~\bibnamefont{Eremeichenkova}},
  \bibinfo{author}{\bibfnamefont{L.}~\bibnamefont{Metlov}}, \bibnamefont{and}
  \bibinfo{author}{\bibfnamefont{A.}~\bibnamefont{Morozov}},
  \eprint{http://arxiv.org/abs/physics/0204040}.

\bibitem[{\citenamefont{Ogilvie and Wang}(1992)}]{b24}
\bibinfo{author}{\bibfnamefont{J.}~\bibnamefont{Ogilvie}} \bibnamefont{and}
  \bibinfo{author}{\bibfnamefont{F.}~\bibnamefont{Wang}}, \bibinfo{journal}{J.\
  Mol.\ Struct.} \textbf{\bibinfo{volume}{273}}, \bibinfo{pages}{277}
  (\bibinfo{year}{1992}).

\bibitem[{\citenamefont{Metlov}()}]{b25}
\bibinfo{author}{\bibfnamefont{L.}~\bibnamefont{Metlov}},
  \eprint{http://arxiv.org/abs/nlin.PS/0204041}.

\end{thebibliography}

\end{document}